\documentclass[12pt]{iopart}

\usepackage{amssymb}
\usepackage{graphicx}
\begin{document}

\def\be{\begin{equation}}
\def\ee{\end{equation}}
\def\de{\Delta}
\def\sig{\Sigma}
\newcommand{\bea}{\begin{eqnarray}}
\newcommand{\eea}{\end{eqnarray}}
\newcommand{\eps}{\varepsilon}
\newcommand{\Vef}{V_{\mbox{\scriptsize eff}}}

\title {Microscopic evaluation of the pairing gap}
% {\bf Microscopic evaluation of the pairing gap}
\author{M Baldo$^1$, U Lombardo$^2$, S S Pankratov$^3$, and E E Saperstein$^3$}

\address{$^1$ INFN, Sezione di Catania, 64 Via S.-Sofia, I-95125 Catania,
Italy}
\ead{baldo@ct.infn.it}
\address{$^2$ INFN-LNS and
University of Catania, 44 Via S.-Sofia, I-95125 Catania, Italy}
\ead{lombardo@lns.infn.it}
\address{$^3$ Kurchatov Institute, 123182, Moscow, Russia}
\ead{pankratov@pretty.mbslab.kiae.ru}
\ead{saper@mbslab.kiae.ru}

%Publishing, Dirac
%House, Temple Back, Bristol BS1 6BE, UK}
%\ead{custserv@iop.org}
\begin{abstract}
We discuss the relevant progress that has been made in the last few years on the microscopic theory of the
pairing correlation in nuclei and the open problems that still must be solved in order to reach a satisfactory
description and understanding of the nuclear pairing. The similarities and differences with the nuclear matter
case are emphasized and described by few illustrative examples. The comparison of calculations of different
groups on the same set of nuclei show, besides agreements, also discrepancies that remain to be clarified. The
role of the many-body correlations, like screening, that go beyond the BCS scheme, is still uncertain and
requires further investigation.

\end{abstract}


\maketitle

\section{Introduction}

Despite the pairing in nuclei was established more than fifty years ago, the underlying interaction processes
that are responsible of this remarkable and important correlation are not yet understood completely. The main
reason of the difficulty is that the microscopic theory of nuclear pairing  must rely on the effective
interaction among nucleons, that is not known from first principles. Furthermore it is strong and therefore it
requires the solution of a complex many-body problem. Even in nuclear matter, where the problem is hoped to be
simplified to some extent, an accurate microscopic theory is not available. Numerical Monte-Carlo "exact"
calculations are present in the literature only for low-density pure neutron matter \cite{MC}. Unfortunately
these Monte-Carlo estimates are not in satisfactory agreement among each others \cite{MC} and in any case they
can hardly elucidate the microscopic mechanisms which are at the basis of the onset of pairing.

\par

Besides the many-body aspects of the problem, at least other two features of the nuclear pairing have to be
mentioned. One is related to the fact that the pairing phenomenon occurs close to the Fermi surface, while the
bare nucleon-nucleon (NN) potential necessarily involves also scattering to high energy (or momentum) due to its
strong hard core component, which is one of the main characteristics of the nuclear interaction. It looks
therefore natural to develop a procedure which removes the high energy states and "renormalize" the interaction
into a region close to the Fermi energy. This can be done in different ways, among which the most commonly used
seems to be the Renormalization Group (RG) Method \cite{RG}. A second feature is the relevance of the single
particle spectrum, not only because the density of states at the Fermi surface plays of course a major role, but
also because the whole single particle spectrum has influence on the effective pairing interaction.

\par

Despite these uncertainties of the microscopic theory of pairing in nuclear matter, there is a commonly accepted
point of view that the gap value $\de$ is quite small at the normal density $\rho_0$, much smaller than typical
values of heavy atomic nuclei. This is an indication that the nuclear surface must play a major role in
establishing the value of the pairing gap in finite nuclei.

\par

In the last few years relevant progresse have been made in the microscopic calculations of pairing gap in nuclei
\cite{milan1,milan2,milan3,Pankr1,Pankr2,Pankr3,Dug1,Dug2}. The main established results is that the bare NN
interaction, renormalized by projecting out the high momenta, is a reasonable starting point that is able to
produce a pairing gap which shows a discrepancy with respect to the experimental value not larger than a factor
2. In view of the great sensitivity of the gap to the effective interaction this result does not appear obvious.
The effective pairing interaction constructed within this renormalization scheme \cite{Baldo_1998} explains
qualitatively also the surface relevance. Indeed, this interaction at the surface can exceed the value inside by
one order of magnitude. Direct effect of the surface enhancement of the gap was presented in
\cite{Baldo_1999,Farine_1999} for semi-infinite nuclear matter and in \cite{Baldo_2003} for a nuclear slab.

\par

The role of the surface is also apparent in explaining the puzzling value of the coherence length extracted from
the pairing gap. In fact, for a pairing gap of the order of 1 MeV, the coherence length is expected to be much
larger than the size of the nucleus. It can be argued that at the surface the pairing gap is larger and the
coherence length shorter, i.e. the size of the Cooper pairs should shrink. This was explicitly shown in a recent
paper by N. Pillet, N. Sandulescu, and P. Schuck \cite{Pillet} on the basis of the HFB approach and the
effective D1S Gogny force. It was shown that Cooper pairs in nuclei preferentially are located in the surface
region, with a small size ($2- 3\;$fm). In \cite{Pankratov_2009}, it was examined to what extent the effect
found in \cite{Pillet} is general and independent of the specific choice of NN force. This investigation was
carried out for a slab of nuclear matter, and the pairing characteristics obtained with  the Gogny force were
compared with those with the realistic Paris and Argonne v$_{18}$ forces. The results obtained with the two
realistic forces agree with each other within an accuracy of about 10\% and agree qualitatively with those of
the Gogny force. In \cite{Pankratov_2010} an analogous analysis, with the Paris potential, was made for the
nucleus $^{120}$Sn which is a standard polygon for examining nuclear pairing. Again the results turned out to be
very close to those of \cite{Pillet}. One of them is shown in figure 1, where the ``probability'' distribution
of Cooper pairs,
 \be p(R) = 4\pi R^2 \int\varkappa^2(R,{\bf r})\;d^3r, \label{e:p_R}
\ee is displayed. Here ${\bf R},{\bf r}$ are c.m. and relative coordinates. In Fig. 1 this quantity is
normalized to the total number of Cooper pairs, $N_{\rm Cp}=\int p(R)dR\simeq 10$. One can see that Cooper
pairs, indeed, are strongly concentrated on the surface.

\begin{figure}[]
\centerline{\includegraphics [height=60mm]{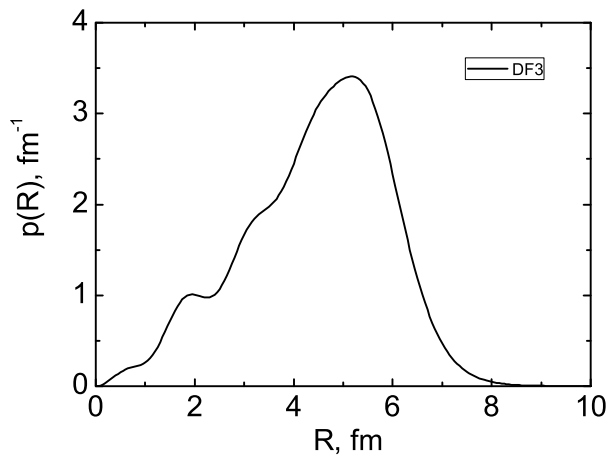}} \caption{
Cooper pair distribution for the $^{120}$Sn nucleus.
 The self-consistent mean field is found with
the Energy Density Functional DF3 of \cite{Fayans} .} \label{e:p(R)DF3}
\end{figure}

\par

Besides the renormalization of the bare interaction of the high
momentum components, other physical effects should be included in
a microscopic approach, like the ones related to the effective
mass, or more generally, the single particle spectrum and the many-body
renormalization of the pairing interaction.  The role of these and
other features of the microscopic scheme are still to be
clarified.

\par

In this paper we will try to summarize the recent achievements and
the main open problems that hinder the
development of an accurate microscopic many-body theory of the nuclear pairing.

\section{Renormalization of the high energy components}

The pairing gap in nuclei is of the order of 1-2 MeV, while the typical energy scale of the bare NN interaction
is of few hundreds MeV. This indicates the difficulty of controlling the microscopic construction of the
effective pairing interaction $\Vef$ at the Fermi surface. In the BCS approach a simplified estimate is commonly
used, the so called weak coupling limit. In this approximation the relationship between the gap $\Delta_{\rm F}$
at the Fermi energy and the effective interaction has the form \be \Delta_{\rm F} \, \approx\, 2 \eps_{\rm F}
\exp (1/\nu_{\rm F}\Vef )\;, \label{e:expf} \ee \noindent where $\eps_{\rm F}$ is the Fermi energy and $\nu_{\rm
F}=m^*k_{\rm F}/\pi^2$ is the density of state at the Fermi energy. The exponential dependence makes the problem
particularly delicate, since a small change in the effective interaction can result in substantial change of the
pairing gap. In phenomenological theories, like the Finite Fermi System (FFS) theory or the
Hartree-Fock-Bogolyubov method with effective forces, the value of $\Vef$ is considered as a phenomenological
parameter (or a set of parameters) to be used to fit the data. Despite the undoubt success of the
phenomenological approaches, the challenge of {\it ab initio} evaluation of the pairing gap remains one of the
most fundamental and not completely solved problems in nuclear physics.

\par

Let us first consider the problem  of reducing the interaction to an effective one close to the Fermi energy. In
general language this can be seen as a typical case that can be approached by an "Effective Theory", where the
low energy phenomena are decoupled from the high energy components. In this procedure the resulting low energy
effective interaction is expected to be independent of the particular form of the high energy components.
However the procedure is not unique. The RG method has been particularly developed for the reduction of the
general NN interaction  keeping the deuteron properties and NN phase shifts up to the energy where they are well
established. In this way a potential, phase equivalent to a known realistic NN potential, can be obtained, which
contains only momenta up to a certain cut-off. The extension of this procedure to the many-body problem appears
in general to require the introduction of strong three-body forces. It has been applied also to the pairing
problem. To illustrate the difficulty of decoupling low and high momentum components for the pairing problem, we
consider the simple case of nuclear matter in the BCS approximation. The BCS equation for the gap $\de(k)$ can
be written as \be
 \de(k) = - \sum_{k'}
 { V(k,k') \over 2 \sqrt{ \eps(k')^2 + \de(k')^2} } \de(k') \:,
\label{e:gapr} \ee \noindent where $V$ is the free NN potential, $\eps(k) = e(k) -\mu$, $e(k)$ is the single
particle spectrum and $\mu$, the chemical potential. It is possible to project out the momenta larger than a
cutoff $k_c$ by introducing the interaction $\Vef(k,k')$, which is restricted to momenta $k < k_c$. It satisfies
the integral equation \be
 \Vef(k,k') = V(k,k')
  - \!\!\sum_{k'' > k_c}
 \!\! \frac { V(k,k'') \Vef(k'',k')} {2E_{k''} }
 \:,
\label{e:vl1} \ee
 \noindent where
 $E(k) = \sqrt{ \varepsilon(k)^2 +
\de(k)^2}$. The gap equation, restricted to momenta $k < k_c$ and with the original interaction $V$ replaced by
$\Vef$, is exactly equivalent to the original gap equation. The relevance of this equation is that for a not too
small cut-off the gap $\de(k)$ can be neglected in $E(k)$ to a very good approximation and the effective
interaction $\Vef$ depends only on the normal single particle spectrum above the cutoff $k_c$. In the RG
approach the low momenta interaction $V_{low-k}(k,k')$ is constructed in such a way to keep the  half-on-shell
two-body T-matrix in free space and therefore the phase shifts. To the extent that the pairing gap is determined
only by the phase shifts, the two approaches are therefore equivalent.

\section{The single particle spectrum and the effective mass}

The pairing gap value is strongly affected by the density of state at the Fermi level, as  the weak coupling
limit (\ref{e:expf}) indicates. In turn, the density of state is proportional to the effective mass. As it is
well known, one has to distinguish between the so-called k-mass, which is generated by the momentum dependence
of the single particle potential, and the e-mass, which is generated by the energy dependence of the single
particle self-energy. The total effective mass is just the product of these two effective masses. The inclusion
of an energy dependent single particle self-energy produces also the so-called Z-factor, i.e. the quasi-particle
strength.

\begin{figure}
\centerline{\includegraphics [height=80mm,width=120mm]{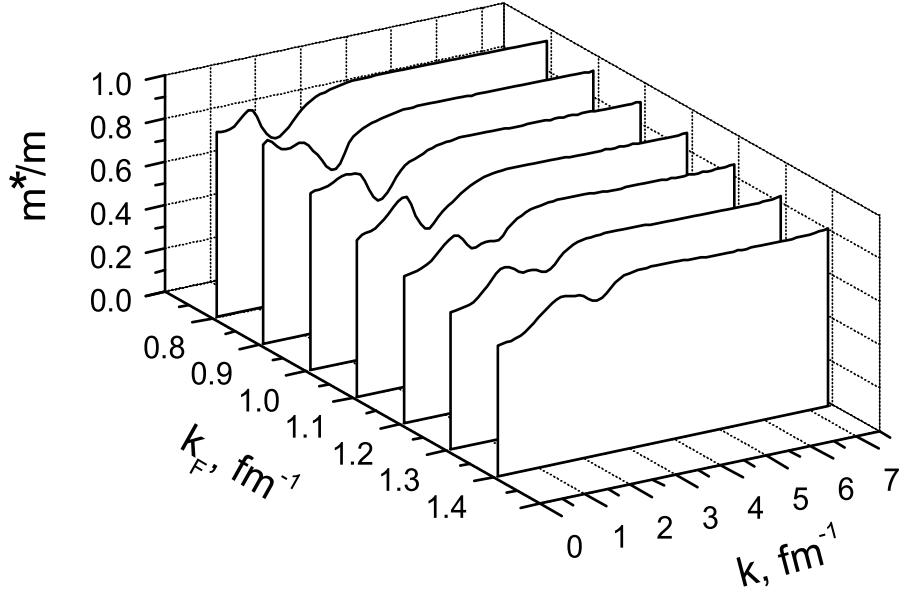}} \vspace{2mm} \caption{Momentum
dependence of the
effective mass $m^*(k)$ at different Fermi momentum values $k_{\rm F}$.}\label{fig:mstk}
\end{figure}

In nuclear matter the four-dimensional gap equation incorporating the momentum and energy dependent irreducible
particle-particle vertex ${\cal V}(k,\eps;k',\eps')$ and the self-energy $\Sigma(k,\eps)$, reads
\cite{pols,books,bg}
 \be
 \de(k,\eps) = -\int\!\frac{d^3 k'}{(2\pi)^3}
 \int\!\frac{d\eps'}{2\pi i}
 \,{\cal V}(k,\eps;k',\eps') {\de(k',\eps') \over D(k',\eps')}\;,
\label{e:gge} \ee with \be \hskip -1cm D(k,\eps) = ({\cal G}{\cal
G}^s)^{-1}=[M(k,+\eps)-\eps-i 0]
 [M(k,-\eps)+\eps-i 0] + \de^2(k,\eps)
\label{e:d} \ee and \be
 M(k,\eps) = {k^2\over 2m} + \sig(k,\mu+\eps) - \mu \:,
\label{e:m} \ee where we define for convenience the energy $\eps$ relative to the chemical potential $\mu$.
${\cal G}^s$ and ${\cal G}$ in (\ref{e:d}) are, correspondingly, the single particle Green functions with and
without pairing.
For realistic systems, ${\cal V}$ and $\Sigma$ cannot be determined in exact way and significant approximations
have to be performed. The usual BCS approximation amounts to replacing the interaction vertex by the (energy
independent) bare nucleon-nucleon potential $V$, and the nucleon self-energy by some realistic s.p.~spectrum.
The latter is characterized by the single particle potential, that can have, in nuclear matter, a momentum
dependence which persists even at high momenta. To illustrate this point we have reported in Fig. \ref{fig:mstk}
the effective mass obtained within the BHF approximation as a function of momentum at different Fermi momenta.
The effective mass has usually a peak close to the Fermi momentum, but it is substantially different from the
bare one even at high momenta.

 As shown in detail in Ref.~\cite{bg}, energy dependence of
the self-energy can be taken into account explicitly provided the
interaction vertex remains static, ${\cal V}(k,\eps;k',\eps') =
\tilde V(k,k')$. In this case, the general gap equation
(\ref{e:d}) is reduced to the  form
\be
 \de(k) = - \sum_{k'}
 { \tilde V(k,k') Z(k') \over 2 \sqrt{ M_s(k')^2 + \de(k')^2} } \de(k') \:,
\label{e:gengapr} \ee  which reminds the BCS gap equation
(\ref{e:gapr}), with the ``symmetrized'' s.p.~energy \be
 M_s(k) \equiv \mathrm{Re}\left({M(k,+e_k)+M(k,-e_k) \over 2}\right)
% \ ,\ e_k = M(k,+e_k)
\label{e:ms} \ee appearing in the denominator and
%an interaction
with the kernel  modified by the spectral factor \be
 Z(k) \equiv
% \left| M_s(k) \right|\,
 \sqrt{M_s(k)^2 + \de(k)^2}\;
 {2\over\pi}\! \int_0^\infty\!\!\! d\eps\,
 \mathrm{Im}{\left(1 \over D(k,\eps)\right) } \:.
\label{e:z} \ee

\begin{figure}
\centerline{\includegraphics
[height=100mm,width=120mm]{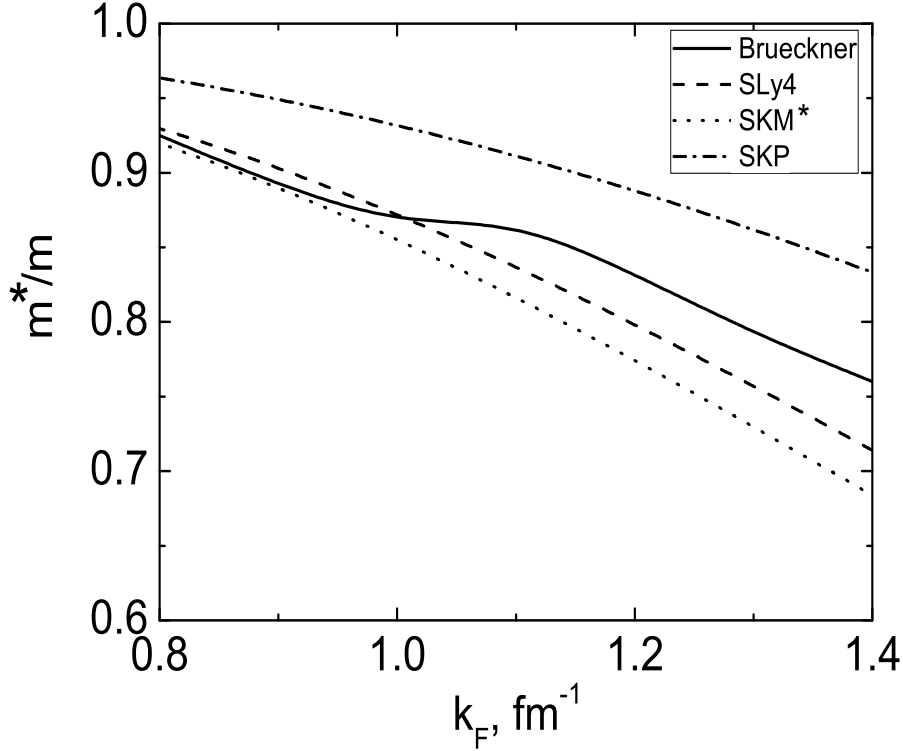}}\vspace{2mm} \caption{
Effective mass $m^*_{\rm F}$ at the Fermi surface depending on the
Fermi momentum $k_{\rm F}$.}\label{fig:mstkF}
\end{figure}

\begin{table}
\caption{\label{tab0} Comparison of the neutron effective mass in the symmetric and asymmetric nuclear matter
from the BHF calculation with that of different kinds of the Skyrme force.}
\begin{indented}
\bigskip

\item[]\begin{tabular}{|c|c|c|c|c|} \hline
                           & SKP \cite{Dob}  & SKM* \cite{SKMS} & SLy4 \cite{SLy4} & BHF \\
\hline
$(m^*/m)_{\rm snm}$         &0.85  & 0.71  &0.74   & 0.78    \\
$\delta (m^*/m)_{\rm anm}$  &0.22  & 0.12  &-0.04  & $-(0.03\div 0.05)$    \\

\hline
\end{tabular}
\end{indented}
\end{table}

In table \ref{tab0} the neutron effective masses at the Fermi surface $m^*(k_{\rm F})$ of symmetric and
asymmetric, $(N-Z)/A=1/6$, nuclear matter found within the BHF approach at the equilibrium density $\rho_0$ is
compared with those of Skyrme forces. In the second line, the difference $\delta (m^*/m)_{\rm anm}=(m^*/m)_{\rm
anm}-(m^*/m)_{\rm snm}$ is given. As one can see, the  SLy4 effective mass  is rather close to the BHF one at
the Fermi surface. Other two kinds of Skyrme force give absolutely different isotopic asymmetry dependence of
$m^*$. The density dependence of the effective mass in symmetric nuclear matter for BHF and the considered
Skyrme forces is reported in figure \ref{fig:mstkF}. Around saturation the trends are smooth, but at lower
density the BHF effective mass has a behavior that interpolates between different Skyrme forces, and therefore
it cannot be reproduced by a given Skyrme force.
\par

\begin{figure}[]
\centerline{\includegraphics [height=60mm]{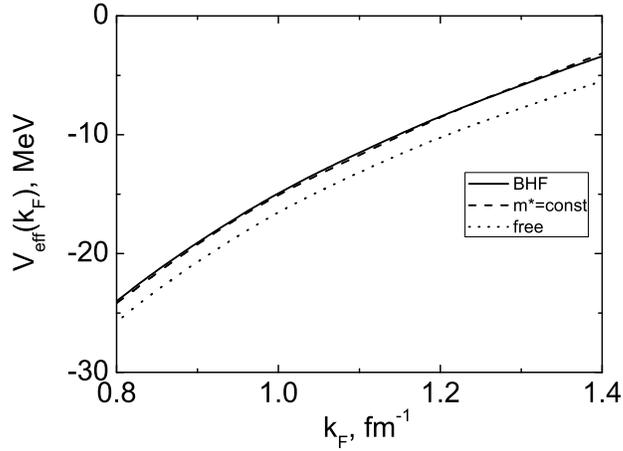}} \caption{
Effective pairing interaction for the free single particle spectrum
(dotted line), the BHF spectrum (full line) and the constant
effective mass spectrum (dashed line).} \label{fig:Veff}
\end{figure}

To illustrate the relevance of the high momenta components $ k > k_c $ on determining the effective pairing
interaction  $V_{\rm eff}$ around the Fermi momentum, we present in figure \ref{fig:Veff} the value of $V_{\rm
eff}$ at the Fermi momentum obtained by solving Eq. (\ref{e:vl1}). We take $k_c = \sqrt{2} k_F$ and consider
three cases : i) the free spectrum, ii) the spectrum obtained from the BHF calculation, and iii) the spectrum
with a constant effective mass, taken at the Fermi momentum and from the BHF results. The calculations are
performed in the density range relevant for the bulk and surface regions of finite nuclei.

\par
\begin{figure}[]
\centerline{\includegraphics [height=60mm]{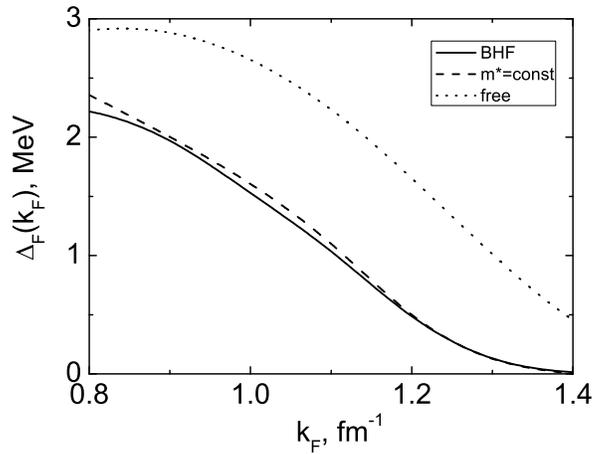}} \caption{
Pairing gap at the Fermi momentum for the three cases of free single
particle spectrum considered in Fig. (\ref{fig:Veff}).}
\label{fig:delta}
\end{figure}

The substantial difference between cases i) and iii) shows the relevance of the high momentum components of the
single particle spectrum. On the other hand, the strong overlap of the curves corresponding to the cases ii) and
iii) suggests that the details of the momentum dependence of the effective mass are less relevant. This will be
helpful for the analysis in finite nuclei, where the approximation of a constant effective mass will be made.
The corresponding pairing gaps at the Fermi momenta are reported in figure \ref{fig:delta}.

\section{The many-body problem}

In atomic nuclei, a new branch of low-laying collective excitations appear, the surface vibrations ("phonons").
They could be interpreted as a Goldstone mode corresponding to spontaneous breaking of translation invariance in
nuclei \cite{KhS}. The ghost dipole $1_1^-$-phonon is the head of this branch. In the self-consistent FFS theory
or other self-consistent approaches, say, the SHF+RPA (or SHFB+qRPA) method, the constraint $\omega_{1_1^-}=0$
is fulfilled identically and the next members of the branch of natural parity states, $2_1^+, 3_1^-$ and so on,
have small excitation energies, less of a typical distance between neighboring shells, $\omega_0 \simeq
\eps_{\rm F}/A^{1/3}$. Up to now, only phenomenological approaches were used to describe the surface vibrations
and their role in the gap equation via the induced interaction. The latter was examined by the Milano group in a
series of papers using various approaches, with the application to the nucleus $^{120}$Sn. In Ref. \cite{milan2}
the study was performed within the Nuclear  Field Theory (see \cite{NFT} and References therein) in the
two-phonon approximation. One and two-phonon terms of the induced interaction were taken into account, as well
as the corresponding corrections to the single-particle states. The latter includes both shifts of the
single-particle energies and spread of their strength which is analogous to  the $Z$-factor in
(\ref{e:gengapr}). The "direct" term of the gap, $\de_{\rm dir}\simeq 0.7\;$MeV, was found in \cite{milan2} with
the Argonne v$_{14}$ NN-potential and the mean field generated by the SLy4 Skyrme force with the effective mass
$m^*\simeq 0.7m$. It is one half of the experimental gap (in \cite{milan2}, which is estimated as $\de_{\rm
exp}\simeq 1.4\;$MeV, while the simplest 3-point formula yields $\de_{\rm exp}^{(\rm 3p)}\simeq 1.3\;$MeV). By
including all the corrections due to low-laying surface vibrations, the complete value $\de=\de_{\rm dir+ ind}$
turns out to be in agreement with the experiment.

In \cite{Barranco_05} the induced interaction itself  was analyzed in detail, again concentrating on low-lying
($\omega_L<5\;$MeV) surface vibrations. The $V_{low-k}$ potential was used as the free NN-interaction, instead
of Argonne v$_{14}$ in \cite{milan2}, and again the experimental value of the gap was obtained as a sum of
practically equal "direct" and "induced" contributions multiplied by the $Z$-factor representing the
quasiparticle strength at the Fermi surface. The latter was estimated on the basis of calculations in
\cite{milan2} as $Z=0.7$.

An essentially different approach was used in \cite{milan3} where all collective states with spin values $J\le
5$ and excitation energies $\omega_J<30\;$MeV were included into the induced interaction term. The surface
vibrations discussed above are only a part of them. Both the natural and unnatural parity states contribute to
the sum. The qRPA method with the SLy4 Skyrme force was used for every $J^{\pi}$-channel. The fragmentation and
self-energy effects were approximately taken into account with the following relation for the total effective
pairing interaction: \be <12|V_{\rm dir + ind}|34> = Z <12|V_{\rm dir} + V_{\rm ind}|34>\,, \label{e:milZ}\ee
where $Z=0.7$ value again was used. A short notation $|12>$ is used in (\ref{e:milZ}) for the two-particle basis
states. To make easier the analysis and comparison with other calculations, separate calculations of the gap
from $V_{\rm dir}$ (Argonne v$_{14}$) and $V_{\rm ind}$, both without the $Z$-factor, were made in
\cite{milan3}. They give $\de_{\rm dir}=1.04\;$MeV and $\de_{\rm ind}=1.11\;$MeV. Using the recipe of Eq.
(\ref{e:milZ}), yields the average gap $\de_{\rm F}=1.47\;$MeV which is again close to the experiment.

 In our opinion, each method used in these papers has some
weak points which could be criticized. In the first case, the contribution of so-called tadpole diagrams
\cite{KhS,Kam_S} was neglected, see also discussion in \cite{Rep}. In the last case, the use of Skyrme force in
the particle-hole spin and spin-isospin channels (unnatural parity states) is questionable. Indeed, the Skyrme
parameters were fitted only to phenomena which are related to "scalar" Landau--Migdal amplitudes, $F,F'$. The
combinations which determine the spin-dependent amplitudes $G,G'$ were not checked up to now by an attempt to
describe corresponding phenomena, e.g. magnetic moments, $M1$-transitions, and so on. This problem is discussed
also in \cite{milan3}, and some complementary calculation was made with $G=G'=0$. It resulted in a very strong
induced interaction yielding very big gap $\de_{\rm F}=2.12\;$MeV. This could be considered as an estimate of
the uncertainty  of such calculations with the use of phenomenological forces. Indeed, even the amplitudes
$F,F'$ which are known sufficiently well in vicinity of the Fermi surface could change significantly when high
energy excitations are considered. Another questionable point is the use in this calculation scheme of the
$Z$-factor found only from the low-lying surface vibrations. The high energy response function included into the
$V_{\rm ind}$ will contribute to the $Z$-factor as well. This contribution comes mainly from the spin-isospin
channel and could be estimated as $Z_{\rm nm}\simeq 0.8$ \cite{Sap_Tol}.

We see that the problem of the screening effect is, indeed, very
difficult, and some approximations must be introduced in the
calculations.  In our opinion, the fact that essentially different
methods were used in \cite{milan2} and \cite{milan3} with
different results for $\de_{\rm ind}$  shows by itself that the
problem of finding contribution of the induced interaction into
the pairing gap in atomic nuclei is far from being solved.

\section{Solution of the ``{\it ab initio}'' gap equation in finite nuclei}

The explicit form of the gap equation (\ref{e:gge}) in the
coordinate representation for a non-uniform system is as follows
\cite{AB}:
%3.1
\bea \lefteqn{ \Delta( {\bf r}_1,{\bf r}_2,\eps) = \int {\cal V}
({\bf r}_1,{\bf r}_2,{\bf r}_3,{\bf r}_4;E=2 \mu, \eps,\eps')
\times}
\nonumber \\
& &{}\times {\cal G}({\bf r}_3,{\bf r}_5,\eps') {\cal G}^s({\bf
r}_4,{\bf r}_6,- \eps') \Delta( {\bf r}_5,{\bf r}_6,\eps') { d\eps'
\over {2 \pi i}} d{\bf r}_3 d{\bf r}_4 d{\bf r}_5 d{\bf r}_6 .
\label{e:delr} \eea As in Sections 2-4, single-particle energies
$\eps,\eps'$ are counted off the chemical potential $\mu$. Dealing
with nuclear matter, we set $\mu = \mu_0 \simeq -16$ MeV (the
leading term in the Weizsaecker mass formula), whereas we have $\mu
\simeq -8$ MeV for stable atomic nuclei, as the $^{120}$Sn nucleus
considered below. As it was discussed above, in this Section we
limit ourselves with the simplest Brueckner-like approach in which
the irreducible vertex ${\cal V}$ coincides with the free
$NN$-potential, ${\cal V}= V$, which is independent of energy. In
this case, the gap $\Delta$ is also independent of energy; hence,
the product of two Green's functions in (\ref{e:delr}) can be
integrated with respect to $\eps'$:
%3.3
\be A^s ({\bf r}_1,{\bf r}_2,{\bf r}_3,{\bf r}_4) = \int { d\eps' \over {2 \pi i}} {\mathcal G} ({\bf r}_1,{\bf
r}_2,\eps' ) {\mathcal G}^s({\bf r}_3,{\bf r}_4,- \eps' ). \label{e:As} \ee The gap equation (\ref{e:delr}) can
be written in a compact form as
%3.4
\be \Delta =  V A^s \Delta. \label{e:delA} \ee Below we deal with the BCS gap equation, not the general one, Eq.
(\ref{e:delr}). This explains why the term {\it ab initio} in the title of the Section is in quotes. In fact, we
speak just about the first step into the problem, i.e. the solution of the BCS-like gap equation with the free
$NN$-potential as the pairing interaction. The mean field potential (or more general, the mass operator) used in
this solution is taken as a phenomenological input. To go beyond the BCS approximation it is necessary, within
an {\it ab initio} method, to calculate, first, the mass operator and, second, corrections to the interaction
block $\mathcal V$ in the gap equation (\ref{e:delr}). The latter includes the induced interaction discussed
above (see also Ref. \cite{Cao}) and three-body forces \cite{Zuo}. It turns out that, even at the level of this
simplest {\it ab initio} calculations, serious contradictions are still present.

The integral equation (\ref{e:delA})  can be reduced to the form adopted in the Bogoliubov method, \be \Delta =
- { V} \varkappa, \label{e:delkap} \ee where the abnormal density matrix $\varkappa= A^s \de$ can be expressed
in terms of $u$ , $v$-functions,
%6
 \be \varkappa({\bf r}_1,{\bf r}_2) = \sum_i u_i({\bf r}_1) v_i({\bf
 r}_2), \label{e:kapuv} \ee which satisfy the system of Bogoliubov equations.
 The summation in (\ref{e:kapuv}) is over the complete set of
 Bogoliubov functions with the eigen energies $E_i>0$.

The Milano group was the first who, in a series of papers \cite{milan1,milan2}, \cite{milan3} and Refs. therein,
solved the gap equation (\ref{e:delkap}) with the realistic Argonne $NN$-force v$_{14}$ for the nucleus
$^{120}$Sn. The latter was chosen for a definite reason. Indeed, the chain of semi-magic tin isotopes is a
traditional polygon for examining nuclear pairing \cite{Dob}, \cite{Fayans}. The nucleus under discussion is in
the middle of the chain and the number of neutrons participating in the pairing, those above the closed shell
$N=50$, is sufficiently big to use the approximation of the "developed" pairing, used, in fact, in
(\ref{e:delkap}). This approximation implies neglecting the particle number fluctuations \cite{Bohr} typical of
the BCS-like theories. The set of Bogoliubov equations was solved directly in the basis \{$\lambda$\} of the
states with a fixed limiting energy $E_{\max}$. Such direct method is difficult because of a slow convergence of
sums over intermediate states $\lambda$ in the gap equation. These sums are analogous to integrals in the
momentum space in the gap equation for infinite nuclear matter, see Section 2. In \cite{milan1} the value of
$E_{\max}{=}600\;$MeV was used, and in \cite{milan2,milan3}, $E_{\max}{=}800\;$MeV. The analysis of
\cite{Pankr2} showed that the use of such big value of $E_{\max}$ permits to find  $\Delta$ only with accuracy
of 10\%. In \cite{milan1} the Shell Model basis was used with the Saxon-Woods potential and the bare mass,
$m^*=m$, and the value $\Delta{=}2.2\;$MeV was obtained which is by a factor one and half greater than the
experimental one (${\simeq} 1.3 \div 1.4\;$MeV). Evidently, this contradiction forced the authors to use in
further works the self-consistent basis of the Skyrme-Hartree-Fock (SHF) method with the density depending
effective mass $m^*(\rho)\neq m$. In particular, the popular Sly4 force was used, which is characterized by a
small effective mass, equal to $m^*\simeq 0.7m$ in nuclear matter at the normal nuclear density. Solving the BCS
gap equation with such a basis the value of $\Delta$ $\simeq 0.7\;$MeV was obtained in \cite{milan2} and $\simeq
1.0\;$MeV in \cite{milan3}.   A close value for the gap in the BCS equation was found in \cite{Pankr2} for the
nuclear slab with parameters which mimic $^{120}$Sn nucleus. This calculation was done with Argonne v$_{18}$
potential which differs only slightly from the v$_{14}$ one, but the single-particle basis with $m^*=m$ was
used. Remind that in \cite{milan2,milan3} corrections to the BCS gap due to induced interaction were considered
bringing the results rather close to the experimental data as discussed in Section 4. The crucial dependence of
the gap on $m^*$, which is easily seen in weak coupling limit (\ref{e:expf}) for the gap in nuclear matter, is,
of course, the main reason of so strong variation of the BCS gap from \cite{milan1} to \cite{milan3}. In finite
nuclei, this dependence is weaker than in nuclear matter, as the surface region plays here the main role and one
has $m^*(\rho)\to m$ in this region. However, the $m^*$ effect remains strong.

Recently, results from the {\it ab initio} BCS equation (\ref{e:delkap}) for a number of semi-magic nuclei were
published \cite{Dug1,Dug2}. They are based on  the  soft realistic low-k force discussed above which was
calculated starting from the Argonne v$_{18}$ potential, with the same self-consistent Sly4 basis, i.e. the same
effective mass, as in \cite{milan3}. For the nucleus $^{120}$Sn under consideration the value $\Delta {\simeq}
1.6\;$MeV was obtained. This value exceeds the experimental one, which raises some questions. Indeed, although
there are discrepancies in absolute value of the corrections to the BCS gap (see for instance \cite{milan3} and
\cite{Av_Kam}), their sign is more or less definite. All
 calculations of these corrections, to our knowledge, increase
 $\Delta$ considerably. Hence the BCS equation has to lead to the
 gap value smaller  than the experimental one.
In addition, there is a direct contradiction between the results of the Milano group and the ones of Duguet with
co-authors, despite the BCS problem was solved with very similar inputs. Evidently, there is some difference in
the method of including the effective mass, which is hidden in \cite{Dug1,Dug2} under the renormalization  of
v$_{18}$ into $V_{low-k}$.

In \cite{Pankratov_2009} an attempt was taken to clarify the reasons of this contradiction. The BCS gap equation
(\ref{e:delkap}) was solved for the same $^{120}$Sn nucleus with the separable form of Paris potential, which
simplifies calculations. Our experience of calculations for nuclear slab \cite{Pankr1,Pankr2} shows that the
difference between the Paris potential and the Argonne v$_{18}$ one for the gap value is of the order of 0.1 MeV
which is significantly smaller than the deviation under discussion. For projecting out the high momenta
contribution, the so-called Local Potential Approximation (LPA) method was used. This new version of the local
approximation was introduced by our group for semi-infinite nuclear matter and nuclear slab system, see the
reviews \cite{Rep,ST}. In general, this method is analogous to the renormalization scheme for solving the gap
equation in infinite nuclear matter described in Section 2. To solve the gap equation in the form (\ref{e:delA})
for finite systems, we split the complete Hilbert space $S$ of the pairing problem  into the model subspace
$S_0$, including the single-particle states with energies less than a fixed value $E_0$, and the subsidiary one,
$S'$. Correspondingly, the two-particle propagator (\ref{e:As}) is the sum \be A^s=A_0^s+A'\,.\label{e:Asum} \ee
The notation becomes obvious if one expands (\ref{e:As}) in the basis of single-particle functions
$\phi_{\lambda}({\bf r})$, \be A^s ({\bf r}_1,{\bf r}_2,{\bf r}_3,{\bf r}_4) = \sum_{\lambda_1 \lambda_2\lambda
_3 \lambda_4} A^s_{\lambda_1 \lambda_2\lambda _3 \lambda_4}\; \phi^*_{\lambda_1}({\bf
r}_1)\phi^*_{\lambda_2}({\bf r}_2)\phi_{\lambda_3}({\bf r}_3)\phi_{\lambda_4}({\bf r}_4)\,. \label{e:Alam}\ee
The model space propagator $A^s_0$ includes the terms of the sum (\ref{e:Alam}) with single-particle energies
$\eps_{\lambda}<E_0$, $A'$ being the remaining part.

The gap equation is reduced to the one in the model space, \be \Delta = \Vef A^s_0 \Delta\,, \label{e:del0}\ee
with the effective pairing interaction obeying the integral equation in the subsidiary space, \be \Vef =
V+VA'\Vef \,. \label{e:vef}\ee The LPA method concerns the solution of (\ref{e:vef}). Solving this equation
directly in coordinate space is rather complicated, not much simpler than
 the gap equation (\ref{e:delA}) in the complete Hilbert space.
The problem could be simplified with the use of the LPA. It turned out that, with very high accuracy, for each
point ${\bf R}$, one can use the formulas for the infinite system in the potential field $U({\bf R})$ (it
explains the term LPA). This simplifies equation for $\Vef$ significantly, in comparison with the initial
equation for $\Delta$. As a consequence, the subspace $S'$ could be chosen as large as necessary. Validity of
the LPA could be justified by finding that, beginning from some value of $E_0$, the result for the gap doesn't
change under additional increase of $E_0$. It turns out that for a nuclear slab, the value of $E_0{\simeq}
20{\div} 30\;$MeV is sufficient \cite{Pankr1,Pankr2}. For finite nuclei, it should be taken a little bigger,
$E_0= 40\;$MeV \cite{Pankr3}. In the slab calculations, we used the bare nucleon mass, $m^*=m$.In this case, the
reduction of high momenta components of the NN-force in (\ref{e:vef}) is, in fact, quite similar to the
renormalization procedure resulting in a low-k interaction \cite{low-k}. The difference is that in the LPA the
renormalization is coordinate dependent.

\begin{figure}
\centerline{\includegraphics [height=100mm,width=120mm]
{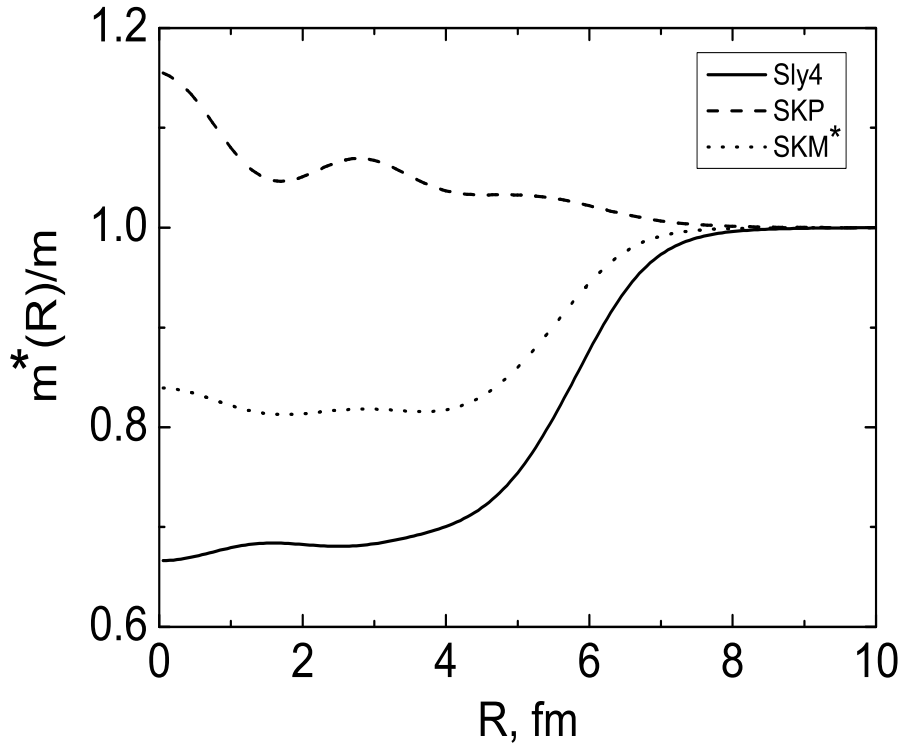}}\vspace{2mm} \caption{Coordinate dependence of
the effective mass for different Skyrme forces in the $^{120}$Sn
nucleus.}\label{fig:mstSk}
\end{figure}

The gap equation in the model space $S_0$ was solved in the $\lambda$-representation with the use of different
single-particle bases   $\phi_{\lambda}$. A discretization method of the continuum spectrum was used with the
wall radius $L{=}16\;$fm. Increasing the radius to $L{=}24\;$fm does not practically influence the results. The
radial eigen-functions  $R_{nlj}(r)$ were found with a step $h{=}0.05\;$fm. We have used the Shell-Model basis
with the Saxon--Woods potential with a standard set of parameters and also with several self-consistent basis
obtained with different methods : the Generalized Energy Density Functional method by Fayans et al.
\cite{Fayans} with the functional DF3, and the SHF method as well with different kinds of Skyrme forces. The
bare mass, $m^*{=}m$, is used in the first method, just as in the Shell Model, whereas in the SHF method the
effective mass is not equal to  $m$ and is density dependent. To clarify the role of the effective mass, we
chose two kinds of the Skyrme force, SKP and SKM*, for which the difference between $m^*$ and $m$ is quite
moderate, and the popular SLy4 force, with significant difference of $m^*$ from $m$, see Fig. \ref{fig:mstSk}.
We see that the SKP effective mass deviates significantly from that for symmetric nuclear matter, Fig.
\ref{fig:mstkF}. This is due to the strong isospin asymmetry effect ($(N-Z)/A=1/6$ for $^{120}$Sn nucleus) for
this kind of Skyrme force. For SLy4 and SKM* this effect is rather modest. Remind that  the Sly4 basis was used
in calculations of $\Delta$ by Milano group and by Duguet with coauthors as well. As to the calculation of the
effective interaction in the subspace $S'$, we first put $m^*{=}m$.  In table 2 the diagonal matrix elements
$\Delta_{\lambda \lambda}$ of the neutron gap in the $^{120}$Sn nucleus, for 5 levels nearby the Fermi level,
are given for each basis under consideration. The quantity $\Delta_{\rm F}$ is the corresponding Fermi-average
value: $\Delta_{\rm F}=\sum_{\lambda}{(2j+1)\Delta_{\lambda \lambda}}/\sum_{\lambda}(2j+1)$.  As we see, in all
cases except of the last one the found value of the gap exceeds the experimental one,  1.4 MeV, significantly.
This indicates the necessity to take into account the difference between the effective and bare masses in the
{\it ab initio} BCS gap equation.

\begin{table}
\caption{\label{tab1} Diagonal matrix elements $\Delta_{\lambda
\lambda}$ (MeV) with the Paris potential for several kinds of the
self-consistent basis.}
\begin{indented}
\bigskip

\item[]\begin{tabular}{|c|c|c|c|c|c|} \hline
 $ \lambda$ & SW   & DF3  & SKP  & SKM* & SLy4
\\
\hline
3$s_{1/2}$  & 1.52 & 1.64 & 1.55 & 1.55 & 1.17  \\
2$d_{5/2}$  & 1.60 & 1.73 & 1.68 & 1.64 & 1.24   \\
2$d_{3/2}$  & 1.64 & 1.80 & 1.71 & 1.68 & 1.26   \\
1$g_{7/2}$  & 1.85 & 2.11 & 2.02 & 1.91 & 1.37   \\
2$h_{11/2}$ & 1.58 & 1.79 & 1.69 & 1.64 & 1.18   \\
\hline
$\Delta_{\rm F}$       & 1.65 & 1.85 & 1.76 & 1.71 & 1.25  \\
\hline
\end{tabular}
\end{indented}
\end{table}

\begin{figure}[]
\centerline{\includegraphics [height=100mm]{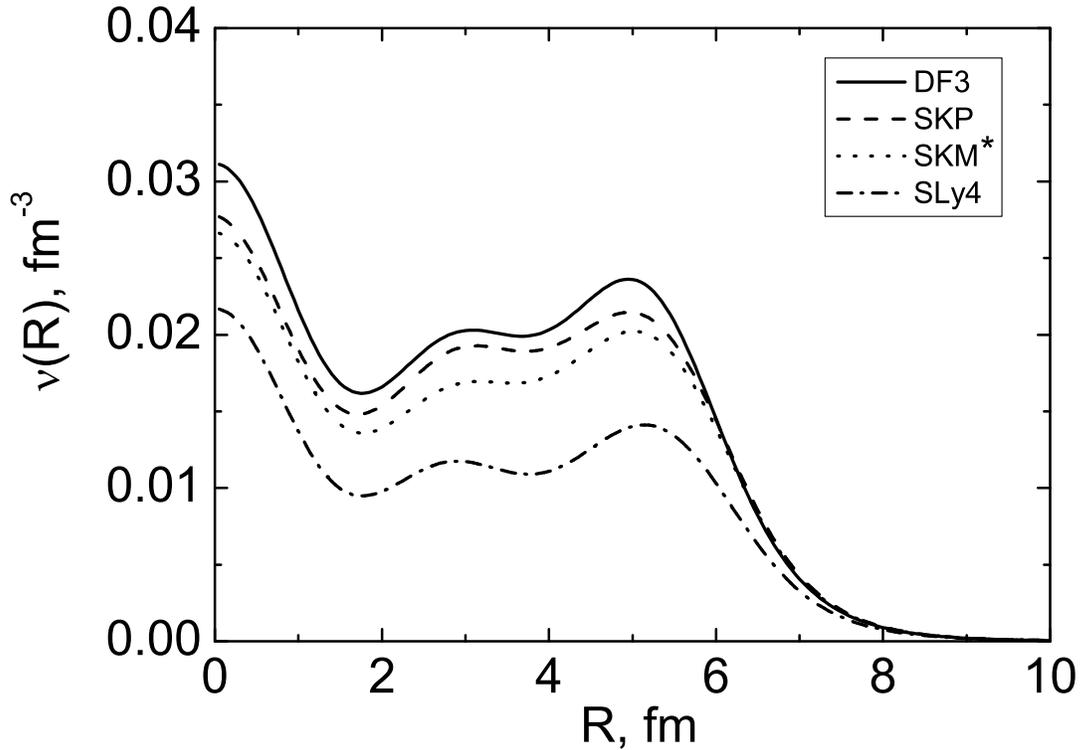}} \caption{
Anomalous density for the $^{120}$Sn nucleus calculated with
different self-consistent mean fields.} \label{fig:nu(R)}
\end{figure}

In Fig. \ref{fig:nu(R)}, for each kind of mean field, the anomalous density $\nu(R)=\varkappa(R,r=0)$ is
displayed. Here the notation ${\bf R} =({\bf r}_1+{\bf r}_2)/2, {\bf r} ={\bf r}_2-{\bf r}_1$ is used. As the
gap $\Delta$ is proportional to this quantity,
  $\nu(R)$  depends on the effective mass in the same way as the matrix elements
  $\Delta_{\lambda \lambda}$ in table 2. Indeed, the anomalous densities
  for the effective forces
 DF3, SKP and SKM* are rather close to each other, whereas for the  SLy4 force
 with a small effective mass the anomalous density is suppressed significantly.
 There is a pronounced surface maximum of $\nu(R)$ for all the cases.
 This leads to the surface enhancement of the pairing gap found in
 \cite{Baldo_2003} for the slab and in  \cite{Pillet}, for spherical nuclei.

Until now, in the case of the Skyrme forces, we took into account the difference between the effective and bare
masses only inside the model space, whereas we put $m^*{=}m$ for calculating $\Vef$. At this point, there is a
principal difference between our
 LPA method and calculations of
\cite{milan2,milan3} and \cite{Dug1,Dug2}, where the SLy4 effective mass was used for all the $\lambda$-states.
For a closer comparison, we made a modification of the LPA method, which permits to take into account the
density dependent effective mass $m^*_n(\rho_n,\rho_p)$ for a part of the space  $S'$, including momenta
$k<\Lambda$, where $\Lambda$ is a parameter. Following the idea of LPA, with the potentials $U_n( R),U_p( R)$ ,
it is natural to determine at each point  $ R$  the densities of each type $\tau=n,p$ of nucleons with
quasi-classical formulas:
 $\rho_{\tau}(R)=(k_{\rm F}^{\tau}(R))^3/3\pi^2$,
$k_{\rm F}^{\tau}(R)=[2m^*_{\tau}(\rho_n(R),\rho_p(R))(\mu_{\tau}-U_{\tau}(R))]^{1/2}$, where $\mu_n,\mu_p$ are
the chemical potentials of neutrons and protons in the nucleus under consideration. For the functional  SLy4 we
made several alternative calculations with different values of $\Lambda$. They are denoted as  SLy4-1
($\Lambda{=}3\;$fm$^{-1}$), SLy4-2 ($\Lambda{=}4\;$fm$^{-1}$) and SLy4-3 ($\Lambda{=}6.2\;$fm$^{-1}$). The first
two versions mimic calculations of \cite{Dug1,Dug2}, the latter, of \cite{milan2,milan3}. The obtained gap
values are given in table 3.

\begin{table}
\caption{\label{tab2} Diagonal matrix elements $\Delta_{\lambda
\lambda}$ with the Paris potential for the Sly4 basis depending on
the way of account for the effective mass  in equation for
$\Vef$.}
\bigskip
\begin{indented}
\item[]\begin{tabular}{|c|c|c|c|c|}
\hline
 $\lambda $ &  SLy4 & Sly4-1 & Sly4-2 & Sly4-3\\
\hline
3$s_{1/2}$  & 1.17 & 1.07 &0.88 & 0.76 \\
2$d_{5/2}$  & 1.24 & 1.13 &0.93 & 0.80 \\
2$d_{3/2}$  & 1.26 & 1.15 &0.95 & 0.83 \\
1$g_{7/2}$  & 1.37 & 1.23 &0.99 & 0.85 \\
2$h_{11/2}$ & 1.18 & 1.08 &0.88 & 0.75 \\
\hline
$\Delta_{\rm F}$& 1.25 & 1.14 & 0.92 & 0.80\\

\hline
\end{tabular}
\end{indented}
\end{table}

\begin{table}
\caption{\label{tab3}  The same as in table \ref{tab2}, but for
the Argonne v$_{18}$ potential.}
\bigskip
\begin{indented}
\item[]\begin{tabular}{|c|c|c|c|c|}
\hline
 $\lambda $ &  SLy4 & Sly4-1 & Sly4-2 &Sly4-3 \\
\hline
3$s_{1/2}$   &1.23 &1.10 &0.83& 0.56\\
2$d_{5/2}$   &1.32 &1.18 &0.89& 0.61\\
2$d_{3/2}$   &1.34 &1.20 &0.92& 0.63\\
1$g_{7/2}$   &1.48 &1.31 &0.96& 0.64\\
2$h_{11/2}$  &1.27 &1.13 &0.85& 0.57\\
\hline
$\Delta_{\rm F}$&1.34&1.19 &0.89& 0.60\\

\hline
\end{tabular}
\end{indented}
\end{table}

Let us now repeat calculations with the SLy4 basis, but for Argonne force v$_{18}$. Results are given in table
4. Comparison with table 3 shows that the difference between gap values for the Paris and Argonne force is of
the same order ($\simeq 0.1\;$MeV) as for the slab calculations \cite{Pankr2}, with the exception of the SLy4-3
version with the cutoff for $m^*$  in the equation for $\Vef$ equal to $\Lambda_3=6.2\;$fm$^{-1}$. In this case,
the difference is $\simeq 0.2\;$MeV. It is worth of noticing that the sign of the difference changes depending
on $\Lambda$. Namely the Argonne gap exceeds the Paris one for SLy4 ($\Lambda=0$) and SLy4-1
($\Lambda_1=3\;$fm$^{-1}$) versions, but becomes smaller for SLy4-2 ($\Lambda_2=4\;$fm$^{-1}$) and SLy4-3 runs.
Such behavior is qualitatively clear.Indeed, the Paris potential is much harder of the Argonne one, therefore
the relative contribution to $\Vef$ of the momentum region, say, between $\Lambda_2$ and $\Lambda_3$, is less
than for the Argonne potential. Therefore, for the gap equation itself, the role of the corresponding
suppression of $\Vef$ due to putting $m^*<m$ is less in the Paris case than in the Argonne one.

It is rather difficult to make a direct comparison  of table 4 with results of  \cite{Dug1}. In the first
column, the effective mass $m^*\neq m$ is introduced only in the model space, for $\eps_{\lambda}<E_0=40\;$MeV.
In the free space, outside the nucleus, it corresponds to the momentum cutoff $\Lambda=1.4\;$fm$^{-1}$; inside
the nucleus we have $\Lambda\simeq 2\;$fm$^{-1}$.We see that the "average" value of $\Lambda$ is less a bit of
$\Lambda=2\;$fm$^{-1}$ in \cite{Dug1}. In the second column (SLy4-1) we deal with $\Lambda=3\;$fm$^{-1}$. Thus,
we should attribute to the gap value of \cite{Dug1} ($\de \simeq 1.6\;$MeV) an average value of these two
columns, $\de \simeq 1.25\;$MeV, which is noticeably smaller. For comparison with \cite{milan2,milan3} we should
be guided by the last column (SLy4-3), as the corresponding value $\Lambda=6.2\;$fm$^{-1}$ was chosen to
reproduce $E_{\rm max}=800\;$MeV from these calculations. Again, this correspondence is not literal as it takes
place only outside the nucleus where we have $m^*=m$. Inside, due to $m^*\neq m$, the same value of
 $E_{\rm max}$ should correspond to smaller value of $\Lambda\simeq 5.5\;$fm$^{-1}$.
 Again we should take a value between those of the 3-rd column and the last one,
 closer to the latter. In any case, it will be closer
to the result of \cite{milan2} ($\simeq 0.7\;$MeV) than of
\cite{milan3} ($\simeq 1\;$MeV).

The main observation, common to table 3 (Paris) and table 4 (Argonne), is a drastic dependence of the gap on the
$m^*(k)$ behavior in the subsidiary space. In fact, a set of calculations with different cutoff $\Lambda$
imitates, very roughly, the $k$-dependence of the effective mass.In the case that the asymptotic limit
$m^*(k)\to m$ occurs sufficiently far away, the pairing gap in nuclei does depend on the effect of $m^*\neq m$
at high momenta. It is absolutely lost in calculations of $\de$ with low-k force found for small cutoff values.
Evidently, this is the main reason why gap values found in  \cite{Dug1} by solving the BCS gap equation are so
high, keeping in mind corrections due to the induced interaction.

\begin{table}
\caption{\label{tab4}  The same as in table \ref{tab3}, but for the
SKM* Skyrme force.}
\bigskip
\begin{indented}
\item[]\begin{tabular}{|c|c|c|c|c|}
\hline
 $\lambda $ &  SKM* & SKM*-1 & SKM*-2 &SKM*-3 \\
\hline
3$s_{1/2} $ &  1.58   &1.52 &1.40 &1.29  \\
2$d_{5/2} $ &  1.71   &1.63 &1.49 &1.38  \\
2$d_{3/2} $ &  1.75   &1.68 &1.55 &1.43  \\
1$g_{7/2} $ &  2.01   &1.92 &1.76 &1.62  \\
2$h_{11/2}$ &  1.72   &1.64 &1.51 &1.40  \\
\hline
$\Delta_{\rm F}$&1.78 &1.71 &1.57 &1.44 \\
\hline
\end{tabular}
\end{indented}
\end{table}

To confirm the leading role of the effective mass in the problem, we made a series of analogous calculations for
the SKM* force for which the effective mass in the nucleus under consideration is much closer to m, than for the
SLy4 one, see Fig. \ref{fig:mstSk}. The results are given in table \ref{tab4} with the notation similar to that
of table \ref{tab3}. We see that even for the SKM*-3 version with $\Lambda=6.2\;$fm$^{-1}$ the result does not
leave any room for contribution of the induced interaction.

\section{Discussion and conclusions}
In this paper we briefly reviewed the status of the microscopic theory of nuclear pairing. Although to date
there is no consistent theory of nuclear matter pairing, progress has been made in developing approximated
methods yielding comparable results on the density dependence of the pairing gap. However, the results cannot be
applied to finite nuclei directly because the nuclear surface plays a leading role in nuclear pairing and the
standard LDA fails at the nuclear surface. As to finite nuclei, some progress also exists especially in solving
the simplest, in fact BCS-type, "{\it ab initio}" gap equation, where the free NN potential is used as the
effective pairing interaction. Even this equation is not completely microscopic as far as the phenomenological
mean field is used. The inherent technical problems were overcome, first, by the Milan group
\cite{milan1,milan2,milan3} and more recently by other teams \cite{Dug1,Dug2} and \cite{Pankratov_2009}. In the
first and the last cases, the nucleus $^{120}$Sn was considered as "testing sample", whereas in \cite{Dug1}
several isotopic and isotonic chains were considered. The Argonne v$_{14}$ force was used in
\cite{milan1,milan2,milan3}, the low-k force with the cutoff $\Lambda=2\;$fm$^{-1}$ in \cite{Dug1}, and the
Paris potential, in \cite{Pankratov_2009}. In this paper we performed calculations analogous to
\cite{Pankratov_2009} for the Argonne v$_{18}$ potential which differ only a bit from the v$_{14}$ one. All the
results differ from the experimental gap in $^{120}$Sn, $\de_{\rm exp}\simeq 1.3 \div 1.4\;$MeV, not more than
by a factor two which shows relevance of the {\it ab initio} BCS gap equation as a starting point for the
microscopic theory of nuclear pairing. However, more detailed comparison shows that there are contradictions
even  at this "first level" of the problem. Indeed, the BCS gap is $\simeq 1\;$MeV in \cite{milan3} and is
$\simeq 1.6\;$MeV in \cite{Dug1}, whereas inputs look quite similar. Namely, both the calculations use the SLy4
Skyrme force with the effective mass $m^*\simeq 0.7 m$; the Argonne v$_{14}$ force is used  in  \cite{milan3}
and the low-k force in \cite{Dug1}, but the latter could be obtained from the first one with the RGM procedure.
The only important difference of the inputs is the size of the momentum space where the effective mass
contributes: $k<k_{\rm max}\simeq 6\;$fm$^{-1}$ in \cite{milan3} and only $k<\Lambda=2\;$fm$^{-1}$ in
\cite{Dug1}. Indeed, the RGM equation is defined for the free NN scattering where the equality $m^*=m$ is
postulated. In fact, we deal with different $k$-dependence of the effective mass. The equality $m^*\simeq 0.7 m$
takes place for all momenta in \cite{milan3} and only for $k<\Lambda$, in \cite{Dug1}. This reason of the
contradictory  results of \cite{milan3} and  \cite{Dug1} was discussed in \cite{Dug2} and was confirmed with the
analysis of \cite{Pankratov_2009} for the Paris force and in this paper, for the Argonne v$_{18}$ force. We use
the so-called LPA method developed by us previously in which high momenta are excluded via some renormalization
procedure which recall the RGM one but is coordinate dependent and permits to introduce in the high momentum
space the effective mass into the equation for the effective pairing interaction. For the same SLy4 basis, we
changed "by hands" the size of the space where the equality $m^*\simeq 0.7 m$ takes place, putting $m^*=m$
outside. Changing this dividing point we evolve from the situation close to that of \cite{Dug1} to the one of
\cite{milan3}, although the correspondence, of course, is not literal. In the first limit, we obtained the value
of $\de_{\rm F}\simeq 1.25\;$MeV which is noticeably smaller than that in \cite{Dug1}. In the second one, we
obtained the result closer to that of \cite{milan2} than of \cite{milan3}. But, in our opinion, fixing some
numerical contradictions is  not of primary importance. This disagreement can be resolved. Much more important
is the very high sensitivity of the pairing gap to the $k$-dependence of the effective mass found in our
analysis. This dependence could be hardly guessed phenomenologically by a lucky Skyrme-type ansatz. Indeed, any
such ansatz deals with a number, not a function! Therefore an additional microscopic ingredient, namely a theory
of the $k$-dependent effective mass, is necessary even at this first level of the problem.

The next open problem is the inclusion of the many-body corrections to the BCS gap equation. The induced pairing
interaction due to exchange by virtual surface vibrations and other particle-hole excitations is the main of
them \cite{milan2,milan3}. Up to now, this problem was studied only within phenomenological approaches and is
far from being solved. Any attempt to attack it from first principles hits the absence of the general
microscopic nuclear theory.

\section{Acknowledgments}
We thank H.-J. Schulze for useful  discussions. Two of the authors
(S.S.P. and E.E.S.) thank INFN (Sezione di Catania) for
hospitality during their stay in Catania. This research was
partially supported by  the Grants of the Russian Ministry for
Science and Education  NSh-3004.2008.2 and  2.1.1/4540, the joint
Grant of RFBR and DFG, Germany,  No. 09-02-91352-NNIO\_a, 436 RUS
113/994/0-1(R), by the RFBR grants 07-02-00553-a, 09-02-01284-a
and 09-02-12168-ofi\_m.

{}

\end{document}